\documentclass[journal]{IEEEtran}

\usepackage[T1]{fontenc}
\usepackage{amsfonts}
\usepackage{epsf}
\usepackage{graphicx}
\usepackage{algorithm}
\usepackage{algpseudocode}
\usepackage{subfigure}
\usepackage{amsmath}
\usepackage{cite}
\usepackage{amssymb}

\newcommand{\norm}[1]{\left\lVert#1\right\rVert}
\allowdisplaybreaks \allowdisplaybreaks[4]
\begin{document}

\title{\LARGE Energy-Efficient Data Collection in UAV Enabled Wireless Sensor Network}

\author{Cheng~Zhan,~\IEEEmembership{Member,~IEEE,}
        Yong~Zeng,~\IEEEmembership{Member,~IEEE,}
        and~Rui~Zhang,~\IEEEmembership{Fellow,~IEEE}

\thanks{C. Zhan is with the School of Computer and Information Science, Southwest University,
Chongqing 400715, China. (e-mail: zhanc@swu.edu.cn).}
\thanks{Y. Zeng and R. Zhang are with the Department of Electrical and Computer
Engineering, National University of Singapore, Singapore 117583 (e-mail:
\{elezeng, elezhang\}@nus.edu.sg).}}

\maketitle
\thispagestyle{empty}

\begin{abstract}
In wireless sensor networks (WSNs), utilizing the unmanned aerial vehicle (UAV) as a mobile data collector for the ground sensor nodes (SNs) is an energy-efficient technique to prolong the network lifetime. Specifically, since the UAV can sequentially move close to each of the SNs when collecting data from them and thus reduce the link distance for saving the SNs' transmission energy. In this letter, considering a general fading channel model for the SN-UAV links, we jointly optimize the SNs' wake-up schedule and UAV's trajectory to minimize the maximum energy consumption of all SNs, while ensuring that the required amount of data is collected reliably from each SN. We formulate our design as a mixed-integer non-convex optimization problem. By applying the successive convex optimization
technique, an efficient iterative algorithm is proposed to find a sub-optimal solution. Numerical results show that the proposed scheme achieves significant network energy saving as compared to benchmark schemes.
\end{abstract}

\vspace{-0.22in}

\IEEEpeerreviewmaketitle

\section{Introduction}
Wireless sensor networks (WSNs) usually constitute
a large number of low-cost sensor nodes (SNs) that are typically powered by fixed energy sources such as battery, which are difficult to be recharged once deployed\cite{DWu}. Therefore, energy-efficient sensing and communication techniques for SNs are crucial to prolong the lifetime of WSNs.

There has been a growing interest recently in employing the unmanned aerial vehicle (UAV) as a mobile data collector for the ground SNs in WSN \cite{AE}. By leveraging its high mobility, UAV is capable
of collecting data from the SNs energy-efficiently, since it can sequentially visit the  SNs and collect data from them only when it moves sufficiently close to each SN. Thus, the link distance from each active SN to the UAV is significantly reduced, which saves the transmission energy of all SNs. It has been shown that short-distance line-of-sight (LoS) communication links between UAV and ground terminals can be efficiently exploited in various UAV-enabled wireless networks for performance enhancement by properly designing the UAV's trajectory\cite{YZeng1,YZeng3}.

For UAV-enabled WSNs, sleep and wake-up mechanism is another useful technique to save the energy consumption of SNs\cite{SSay}. With such a mechanism, the SNs remain in the
sleep state until they receive the waking up beacon signal with good strength
from the nearby UAV, at which time they will wake up and start sending data to the UAV,
while after the transmission it will return to the sleep state. There are two critical issues in designing UAV-enabled WSNs for data collection. The first one is due to the limited battery energy of
SNs. The wake-up schedule of SNs should thus be appropriately designed so that each SN can complete its data transmission with minimum energy consumption. The second issue is due to the highly dynamic wireless channels between the SNs and the moving UAV, which
are prone to packet loss \cite{NAhmed}, especially for the practical case when multi-path induced channel fading is present. Thus, the trajectory of the UAV should be properly designed to ensure that each SN can transmit data with low outage probability when it is in its wake-up state.

The problem of jointly designing the SNs' wake-up schedule and the UAV's trajectory for energy-efficient data collection is new and challenging, which has not been rigorously studied to our best knowledge. The prior work \cite{FJiang} studied the
UAV's trajectory design via heading control for sum-rate maximization of ground users in their uplink communications with the UAV. A cyclical multiple access scheme was also proposed in \cite{JLyu}, for supporting delay-tolerant data transmission from ground terminals to the UAV in a periodic manner. However, the above works did not aim to minimize the user energy consumptions in the UAV's trajectory design. It is worth noting that an optimization framework for energy-efficient UAV-to-ground communication via trajectory design was recently developed in \cite{YZeng}, but only the UAV's energy consumption was considered.

Under a general fading channel model for the SN-UAV links, this letter studies the joint optimization of SNs' wake-up schedule and UAV's trajectory to achieve reliable and energy-efficient data collection in UAV-enabled WSNs. The aim is to minimize the maximum energy consumption of all SNs while ensuring that a target amount of data is collected reliably from each SN. The design is formulated as a mixed-integer non-convex optimization problem, which is difficult to be optimally solved in general. By applying the successive convex optimization
technique, an efficient iterative algorithm is proposed to find a sub-optimal solution for our design. Numerical results show that the proposed scheme achieves significant energy savings for the SNs  as compared to the benchmark schemes with fixed data collector position or simple straight trajectory of the UAV.\vspace{-0.1in}

\section{System Model and Problem Formulation}\label{model}
We consider a WSN where a UAV is employed as a mobile data collector to gather information from $K$ SNs on the ground, which are denoted by $\{u_k, 1\leq k\leq K\}$. The location of $u_k$ is
denoted by $\mathbf{w}_k\in \mathbb{R}^{2\times 1}$. Each SN $u_k$ generates sensing data of size $S_k$ bits, and the UAV is regularly dispatached to collect the sensed data for a duration of $T$ seconds. We assume that the UAV flies at a fixed altitude of $H$ meters and denote its maximum flying speed as $V_{\max}$ in meter/second (m/s). The initial and
final locations of the UAV are assumed to be pre-determined, whose horizontal coordinates are denoted as $\mathbf{q}_0, \mathbf{q}_F \in \mathbb{R}^{2\times 1}$, respectively. We assume that $\norm{\mathbf{q}_F-\mathbf{q}_0}\leq V_{\max}T$ so that there exists at least one feasible trajectory for the UAV to move from $\mathbf{q}_0$ to $\mathbf{q}_F$ within time $T$. The UAV's flying trajectory projected on the ground is denoted as $\mathbf{q}(t)\in \mathbb{R}^{2\times 1}, 0 \leq t \leq T$. For convenience, the time horizon $T$ is discretized into $M$ time slots, i.e., $T = M\delta_t$, where $\delta_t$ denotes the elemental slot length such that the UAV's location is considered
as approximately unchanged by the ground SNs within each time slot even at the maximum speed. To this end, we usually consider $V_{\max}\delta_t\ll H$. Therefore, the UAV's trajectory $\mathbf{q}(t)$ can be approximated by the sequence $\{\mathbf{q}[m], 1\leq m\leq M\}$, where $\mathbf{q}[m]\triangleq \mathbf{q}(m\delta_t)$ denotes the UAV's location at time slot $m$.

We assume that the sleep and wake-up mechanism is employed, and at most one SN can be waked up to communicate with the UAV at each time slot. Denote the wake-up schedule variable as $x_{k}[m]$, where $x_k[m]=1$ if $u_k$ is waked up at time slot $m$, and $x_{k}[m]=0$ otherwise. Thus, we have $\sum_{k=1}^{K}x_{k}[m]\leq 1, \forall m$. If $x_{k}[m]=1$, then $u_k$ transmits data with a constant transmission power $P_k$ and a designed transmission rate $R_{k}[m]$ in bits/second/Hz (bps/Hz).

We assume quasi-static block fading channels for the ground-UAV links, where the channel remains unchanged within each fading block and may change over blocks. Furthermore, the duration of each fading
block is typically much smaller than $\delta_t$. As such, the number of fading blocks in each time slot, denoted as $L$, is much larger than $1$ in practice. Under a general fading channel model, the channel coefficient between the UAV and $u_k$ at the $l$-th fading block of time slot $m$ can be modelled
as $h_k[m,l]=\sqrt{\beta_k[m]}\rho_k[m,l]$, where $\rho_k[m,l]$ is a small-scale fading coefficient and $\beta_k[m]$ accounts for the large-scale channel attenuation that depends only on the
distance between the UAV and $u_k$. Let $d_k[m]$ be the distance between the UAV and $u_k$ at time slot $m$. We thus have
\begin{eqnarray}
\beta_k[m]=\beta_0d_{k}^{-\alpha}[m]=\frac{\beta_0}{(H^2+\norm{\mathbf{q}[m]-\mathbf{w}_k}^2)^{\alpha/2}},
\end{eqnarray}
where $\beta_0$ denotes the reference channel power gain at $d_0=1$m, and $\alpha\geq 2$ is the
path loss exponent. Without loss of generality, for any time slot $m$, $\rho_k[m,l]$ are assumed to be
independent and identically distributed (i.i.d.) random variables with $\mathbb{E}[|\rho_k[m,l]|^2]=1$. We assume that the UAV only knows the locations of the SNs as well as the channel distribution information (CDI), namely the values for $\alpha$ and $\beta_0$ as well as the identical distribution of $|\rho_k[m,l]|^2$. Note that due to the time-varying UAV locations, the distribution of $|h_k[m,l]|^2$ keeps unchanged within each time slot but varies over different time slots. Therefore, the transmission rate $R_k[m]$ by the wake-up SN can be designed adaptively over each time slot based on the UAV's location. Once the flying trajectory $\mathbf{q}[m]$, wake-up schedule $x_k[m]$, and transmission rate $R_k[m]$ of the wake-up SN at each time slot are determined, the UAV will wake up the corresponding SNs along its trajectory, and inform each of them the optimized transmission rate over time slots using the downlink control links.

If $u_k$ is in the wake-up state for
communication at time slot $m$, then for the $l$-th fading block of time slot $m$, the achievable rate in bps/Hz is given by
\begin{eqnarray}
C_k[m,l]=\log_2{\left(1+\frac{|h_k[m,l]|^2P_k}{\sigma^2\Gamma}\right)},
\end{eqnarray}
where $\sigma^2$ is the noise power, $\Gamma>1$ is the SNR gap between the practical modulation schemes and the theoretical Gaussian signaling. The outage probability
between $u_k$ and the UAV at the $l$-th fading block of time slot $m$ is then given by
\begin{eqnarray}
&&\hspace{-12mm}p_k[m,l]=\mathbb{P}(C_k[m,l]<R_k[m])\nonumber\\
&&=\mathbb{P}\left(|\rho_k[m,l]|^2<\frac{\sigma^2\Gamma(2^{R_k[m]}-1)}{\beta_k[m]P_k}\right)\nonumber\\
&&=F\left(\frac{\sigma^2\Gamma(2^{R_k[m]}-1)}{\beta_k[m]P_k}\right)\triangleq p_k^{out}[m],
\end{eqnarray}
where $F(\cdot)$ denotes the identical cumulative distribution function (CDF) of $|\rho_k[m,l]|^2$. Note that during each time slot $m$, $p_k[m,l]$ is identical for different fading blocks $l$, and thus is denoted as $p_k^{out}[m]$, which is a non-decreasing function with respect to $R_k[m]$. Therefore, in order to ensure that the target amount of sensing information of each SN is collected reliably by the UAV, $R_k[m]$ should be chosen such that $p_k^{out}[m]=\epsilon$, where $\epsilon$ denotes the maximum tolerable outage probability. As a result, the transmission rate can be expressed as
\begin{eqnarray}
R_k[m]=\log_2\left({1+\frac{F^{-1}({\epsilon})P_k\beta_0}{\sigma^2\Gamma(H^2+\norm{\mathbf{q}[m]-\mathbf{w}_k}^2)^{\alpha/2}}}\right),
\end{eqnarray}
where $F^{-1}(\cdot)$ is the inverse function of $F(\cdot)$.

Let $\mathbf{X}=\{x_k[m], \forall k,m\}$ and $\mathbf{Q}=\{\mathbf{q}[m], \forall m\}$. Our aim is to jointly optimize the wake-up schedule $\mathbf{X}$ and the UAV's trajectory $\mathbf{Q}$ so as to minimize the maximum energy consumption of all SNs, while ensuring that the target amount of data $S_k$ in bits is collected from $u_k$ reliably (i.e., with maximum outage probability $\epsilon$). Define $D_{\max}\triangleq \delta_t V_{\max}$ in meter, $E_k\triangleq \delta_t P_k$ in Joule, and $r_k\triangleq \frac{S_k}{B\delta_t}$ in bps/Hz, where $B$ denotes the channel bandwidth in Hz; then the problem is formulated as
\begin{eqnarray}
\text{(P1)}: \min\limits_{\mathbf{X},\mathbf{Q},\theta} &&\hspace{-4mm} \theta\nonumber\\
\text{s.t.} &&\hspace{-4mm} \sum_{m=1}^{M}x_{k}[m]E_k\leq \theta, \quad\quad\quad\quad\quad\forall k,\\
&&\hspace{-4mm} \sum_{m=1}^{M}x_{k}[m] R_{k}[m]\geq r_k, \quad\quad\quad \forall k,\\
&&\hspace{-4mm} \sum_{k=1}^{K}x_{k}[m]\leq 1, \quad\quad\quad\quad\quad\quad \forall m,\\
&&\hspace{-4mm} x_{k}[m]\in \{0,1\}, \quad\quad\quad\quad\quad\quad \forall k,m, \\
&&\hspace{-4mm} \norm{\mathbf{q}[m]-\mathbf{q}[m-1]}\leq D_{\max}, \forall m\geq 2,\\
&&\hspace{-4mm} \mathbf{q}[1]=\mathbf{q}_0,\mathbf{q}[M]=\mathbf{q}_F.
\end{eqnarray}

Note that $\theta$ is a slack variable that represents the maximum energy consumption among all SNs, and the constraints (6) ensure that the target amount of data from each SN is collected reliably. The constraints (9) and (10) correspond to the UAV's speed and
initial/final location constraints, respectively.

\section{Proposed Solution}\label{Problem}
Problem (P1) is a mixed-integer non-convex problem, which is difficult to be optimally solved
in general. Therefore, in this letter, we aim to obtain an efficient sub-optimal solution to (P1). To this end, we first relax the
binary constraints in (8) as $0\leq x_k[m] \leq 1$, and then solve the relaxed problem iteratively based on the block coordinate descent technique. To reconstruct the binary wake-up schedule variables, note that there are $LM$ fading blocks in total with the time horizon $T$. If the solution $\mathbf{X}$ of the relaxed problem is not binary, we can allocate $N_k[m]=\lfloor Lx_k[m]\rceil$ fading blocks to SN $u_k$ in any time slot $m$, where $\lfloor x\rceil$ denotes the nearest integer of $x$. With sufficiently large $L$, the gap between $N_k[m]$ and $Lx_k[m]$ is practically negligible. For example, if $K=2, L=100$, and the solution of the relaxed problem is $x_1[m]=0.25, x_2[m]=0.75$ in time slot $m$, then $u_1$ can be waked up for $25$ fading blocks while $u_2$ for $75$ fading blocks, which is the optimal solution of the original problem (P1).

In the following, we focus on solving the relaxed problem. Since the relaxed problem is not jointly convex with respect to $\mathbf{X}$ and $\mathbf{Q}$, we adopt the block coordinate descent technique to solve $\mathbf{X}$ and $\mathbf{Q}$ alternately. First, for any given trajectory $\mathbf{Q}$, the integer-relaxed wake-up schedule solution can be obtained by solving the following standard linear program (LP),
\begin{eqnarray}
\text{(P2)}:\min\limits_{\mathbf{X},\theta}&&\hspace{-4mm} \theta \nonumber\\
\text{s.t.} &&\hspace{-4mm} 0\leq x_{k}[m]\leq 1, \quad\quad \forall k,m, \\
&&\hspace{-4mm}(5),(6),(7).\nonumber
\end{eqnarray}

On the other hand, for any given wake-up schedule $\mathbf{X}$, the UAV's trajectory is optimized to maximize the weighted minimum of the communication throughput of all SNs, where the weight is inversely proportional to $r_k$. Specifically, the problem can be formulated as
\begin{eqnarray}
\text{(P3)}:\max\limits_{\mathbf{Q},\eta}&&\hspace{-4mm}  \eta \nonumber\\
\text{s.t.} &&\hspace{-4mm}\frac{1}{r_k}\sum_{m=1}^{M}x_k[m]R_k[m]\geq \eta, \quad\quad\forall k,\\
&&\hspace{-4mm}(9),(10). \nonumber
\end{eqnarray}

Problem (P3) is still a non-convex optimization problem due to the
non-convex constraints (12). However, an efficient approximate solution can be obtained based on the successive convex optimization technique \cite{YZeng}, which is guaranteed to converge to at least a locally optimal solution.
The main idea is to successively maximize a lower bound
of (P3) at each iteration. Let $\mathbf{Q}^{l}=\{\mathbf{q}^{l}[m], \forall m\}$ denote the given UAV's trajectory in the $l$-th iteration. Similar as \cite{YZeng}, by applying the first-order Taylor expansion, $R_k[m]$ in (4) can be lower-bounded as
\begin{eqnarray}
&&\hspace{-8mm}R_k[m]\geq R_{k,l}^{lb}[m]\triangleq A_{k,l}[m]-I_{k,l}[m]\norm{\mathbf{q}[m]-\mathbf{w}_k}^2\nonumber\\
&&\hspace{22mm} +I_{k,l}[m]\norm{\mathbf{q}^{l}[m]-\mathbf{w}_k}^2,
\end{eqnarray}
\begin{eqnarray}
\hspace{-8mm}\text{where}&&\hspace{-5mm}A_{k,l}[m]=\log_{2}\left(1+\frac{F^{-1}(\epsilon)P_k\beta_0}{\sigma^2\Gamma J_{k,l}[m]^{\alpha/2}}\right),\\
&&\hspace{-10mm}I_{k,l}[m]=\frac{F^{-1}(\epsilon)P_k\beta_0(\alpha/2)\log_{2}{e}}{J_{k,l}[m]\left(\sigma^2\Gamma J_{k,l}[m]^{\alpha/2}+F^{-1}(\epsilon)P_k\beta_0\right)},\\
&&\hspace{-10mm}J_{k,l}[m]=H^2+\norm{\mathbf{q}^{l}[m]-\mathbf{w}_k}^2.
\end{eqnarray}
As a result, the UAV's trajectory can be optimized by solving the following problem,
\begin{eqnarray}
\text{(P4)}:\max\limits_{\mathbf{Q},\eta^{lb}}&&\hspace{-4mm}  \eta^{lb} \nonumber\\
\text{s.t.} &&\hspace{-4mm} \frac{1}{r_k}\sum_{m=1}^{M}x_k[m]R_{k,l}^{lb}[m]\geq \eta^{lb} , \quad\quad\forall k,\\
&&\hspace{-4mm}(9),(10).\nonumber
\end{eqnarray}

Since $R_{k,l}^{lb}[m]$ is a concave quadratic function with respect to $\mathbf{q}[m]$, (P4) is a convex quadratically constrained quadratic program (QCQP), which can be solved efficiently by existing software tools such as CVX \cite{MGrant}. Thus, (P3) can be solved by iteratively optimizing (P4)
with the local point $\mathbf{Q}^{l}$ updated in each iteration, which is summarized in Algorithm 1. \vspace{-0.1in}
\begin{algorithm}[h]
\caption{Successive convex optimization for (P3)}
\begin{algorithmic}[1]
\State Initialize the trajectory as $\mathbf{Q}^{0}$;
\State $l\leftarrow 0$; set tolerance $\kappa>0$;
\Repeat
\State Solve the QCQP problem (P4) for given $\mathbf{Q}^{l}$, and denote the optimal solution as $\mathbf{Q}^{l+1}$;
\State $\mathbf{Q}^{l}\leftarrow \mathbf{Q}^{l+1}$; $l \leftarrow l + 1$;
\Until {The fractional increase of the objective value of (P4) is below $\kappa$.}
\label{code:recentEnd}
\end{algorithmic}
\end{algorithm}\vspace{-0.12in}

Similar as in \cite{YZeng}, the resulting objective
values of (P4) in Algorithm 1 are non-decreasing over the iteration. Thus, Algorithm 1 is guaranteed to converge.

The overall algorithm for the integer-relaxed problem (P1) is obtained by optimizing the wake-up schedule $\mathbf{X}$ and trajectory $\mathbf{Q}$
alternately via solving problem (P2) and (P3)
respectively, in an iterative manner, which is summarized in Algorithm 2. \vspace{-0.1in}
\begin{algorithm}[h]
\caption{Iterative algorithm for relaxed (P1)}
\begin{algorithmic}[1]
\State Initialize the trajectory as $\mathbf{Q}^{0}$;
\State $r\leftarrow 0$; set tolerance $\kappa>0$;
\Repeat
\State Solve (P2) for given $\mathbf{Q}^{r}$ to obtain solution $\mathbf{X}^{r}$;
\State Solve (P3) for given $\{\mathbf{X}^{r},\mathbf{Q}^{r}\}$ with Algorithm 1, and denote the solution as $\mathbf{Q}^{r+1}$;
\State $r \leftarrow r + 1$;
\Until {The fractional decrease of the objective value of (P2) is below $\kappa$.}
\label{code:recentEnd}
\end{algorithmic}
\end{algorithm}\vspace{-0.12in}

Next, we consider the convergence of Algorithm 2. To this end, let $\theta(\mathbf{X},\mathbf{Q})$ and $\eta(\mathbf{X},\mathbf{Q})$ be respectively defined as the objective values of problem (P2) and (P3) with given $\mathbf{X}$ and $\mathbf{Q}$, and denote $\theta^{r}=\theta(\mathbf{X}^r,\mathbf{Q}^r)$, $\Omega_k(\mathbf{X}^r,\mathbf{Q}^{r})=\sum_{i=1}^{M}x_k[m]R_k[m]$ for given $(\mathbf{X}^r,\mathbf{Q}^{r})$. It then follows that
\begin{eqnarray}
&&1\overset{(a)}\leq \min_{k}\frac{1}{r_k}\Omega_k(\mathbf{X}^r,\mathbf{Q}^{r})\overset{(b)}=\eta(\mathbf{X}^r,\mathbf{Q}^{r})\nonumber\\
&&\overset{(c)}\leq \eta(\mathbf{X}^r,\mathbf{Q}^{r+1})\overset{(d)}=\min_{k}\frac{1}{r_k}\Omega_k(\mathbf{X}^r,\mathbf{Q}^{r+1}),
\end{eqnarray}
where ($a$) holds since $(\mathbf{X}^r,\theta^r)$ is a solution of (P2) with given $\mathbf{Q}^{r}$, thus it must satisfy the constraints (6); ($b$) and ($d$) hold due to the definition of the problem (P3); ($c$) is true due to the convergence of Algorithm 1. Note that due to constraints (5), $\theta^{r}$ is only related to $\mathbf{X}^r$. Since $\min_{k}\frac{1}{r_k}\Omega_k(\mathbf{X}^r,\mathbf{Q}^{r+1})\geq 1$ based on (18), $(\mathbf{X}^r,\theta^r)$ is a feasible solution of (P2) with given $\mathbf{Q}^{r+1}$. Thus, $\theta^{r+1}\leq \theta^{r}$ as $(\mathbf{X}^{r+1},\theta^{r+1})$ is an optimal solution of (P2) with given $\mathbf{Q}^{r+1}$. Furthermore, since the objective value of (P2) is lower-bounded by a finite value, Algorithm 2 is guaranteed to converge.

\section{Numerical Results}\label{Problem}
\begin{figure}[ht]
\subfigure[UAV's trajectory]{\begin{minipage}{0.5\linewidth}
{\includegraphics[width=1.8 in]{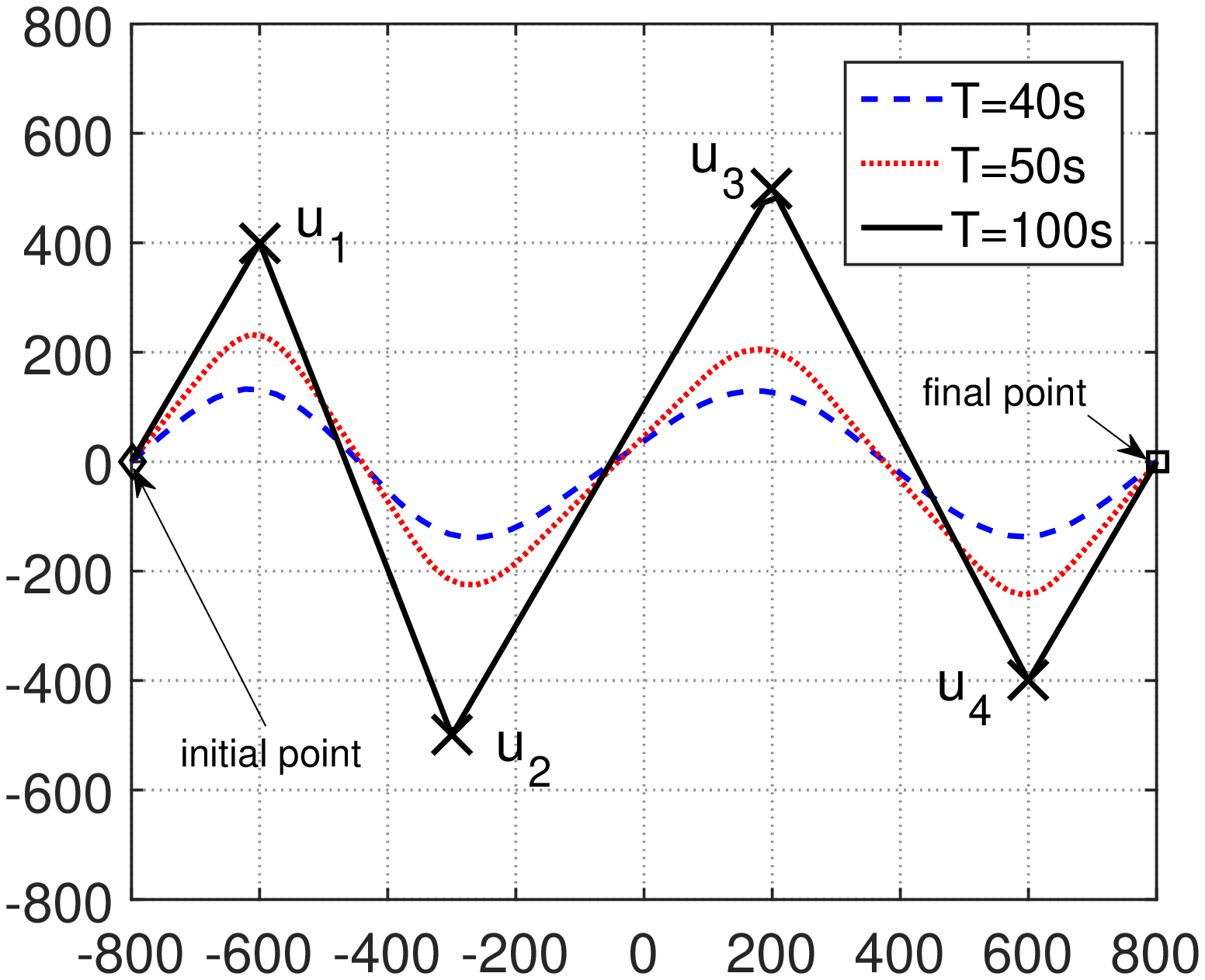}}
\end{minipage}}\subfigure[Wake-up schedule ($T=50s$)]{\begin{minipage}{0.5\linewidth} {\includegraphics[width=1.9
in]{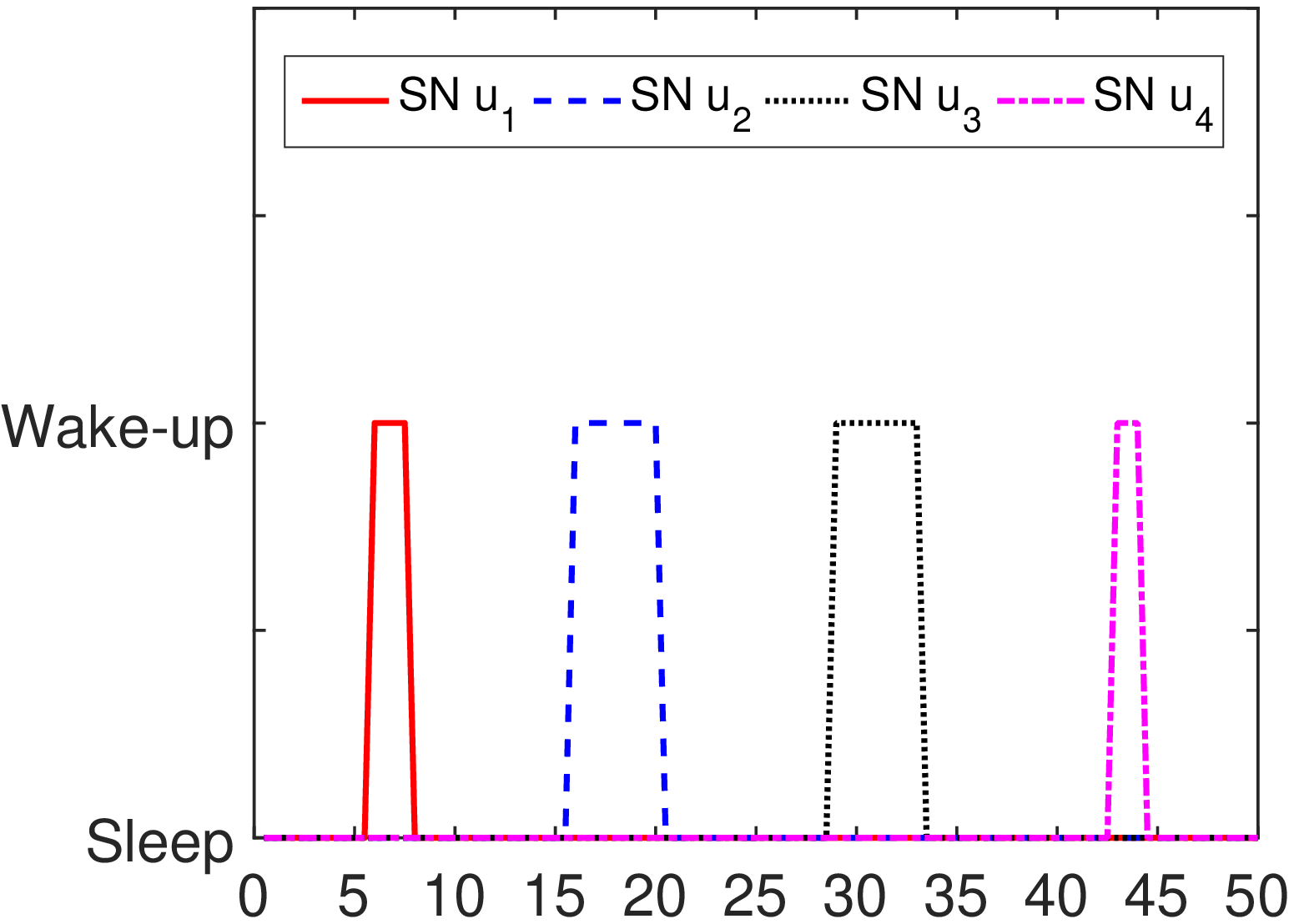}}
\end{minipage}}
\caption{The UAV's trajectory and SNs' wake-up schedule.}
\label{fig:graph1}\vspace{-0.12in}
\end{figure}

In this section, numerical results are provided to verify our proposed design. We consider the practical Rician fading channels with Rician factor $K_c$, and the CDF function $F(\cdot)$ of $|\rho_k[m,l]|^2$ can be expressed as $F(z)=1-Q_1(\sqrt{2K_c}, \sqrt{2(K_c+1)z})$
where $Q_1(a, b)$ is the Marcum-Q function \cite{FOno}. We consider a system with $K=4$ SNs, which are randomly located within an area of size $1.6\times 1.6$ km$^2$. The results obtained are based on one random realization of the SN locations as shown in Fig. \ref{fig:graph1}(a). The UAV's initial and final locations are respectively set as $\mathbf{q}_0=[-800, 0]^T$ and $\mathbf{q}_{F}=[800, 0]^T$ in meter. Furthermore, we set $H=100$m, $V_{\max}=50$m/s, $\delta_t=0.5$s, $B=1$MHz, $\beta_0=-60$dB, $\sigma^2=-110$dBm, $\Gamma=7$dB, $K_c=10$, and $\alpha=2$. The SN's transmission power is set to be $P_k=0.1$W, $\forall k$.

For benchmark comparison, we consider the simple straight flight, where the UAV flies in a straight line from $\mathbf{q}_0$ to $\mathbf{q}_F$ with constant speed $\frac{\norm{\mathbf{q}_F-\mathbf{q}_0}}{T}$. This straight flight is also used as the initial trajectory in Algorithm 2. The optimized trajectories under different $T$ are shown in Fig. \ref{fig:graph1}(a) with $S_k=10$Mbits and $\epsilon=10^{-2}$. It is observed that as $T$ increases, the UAV adjusts its trajectory to move closer to
the SNs. The wake-up schedule of SNs is also shown in Fig. \ref{fig:graph1}(b) for the case of $T=50$s, where it is observed that the SNs remain in sleep states for most of the time and are only waked up when the UAV is moving sufficiently close to them.\vspace{-0.18in}

\begin{figure}
\subfigure[$\theta$ versus $S_k$ ($\epsilon=10^{-2}$)]{\begin{minipage}{0.5\linewidth}
{\includegraphics[width=1.8 in]{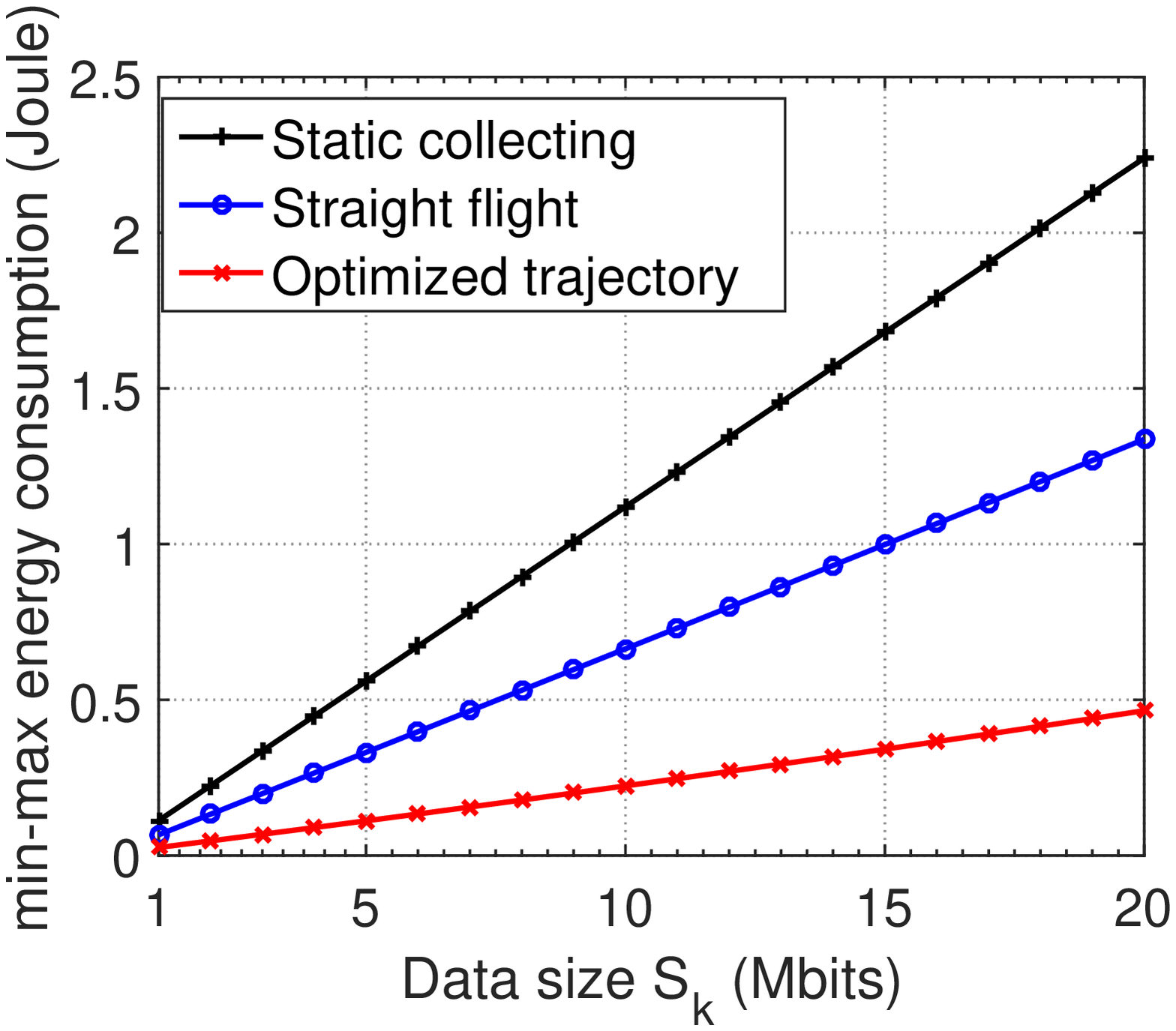}}
\end{minipage}}\subfigure[$\theta$ versus $\epsilon$ ($S_k=10$Mbits)]{\begin{minipage}{0.5\linewidth} {\includegraphics[width=1.8
in]{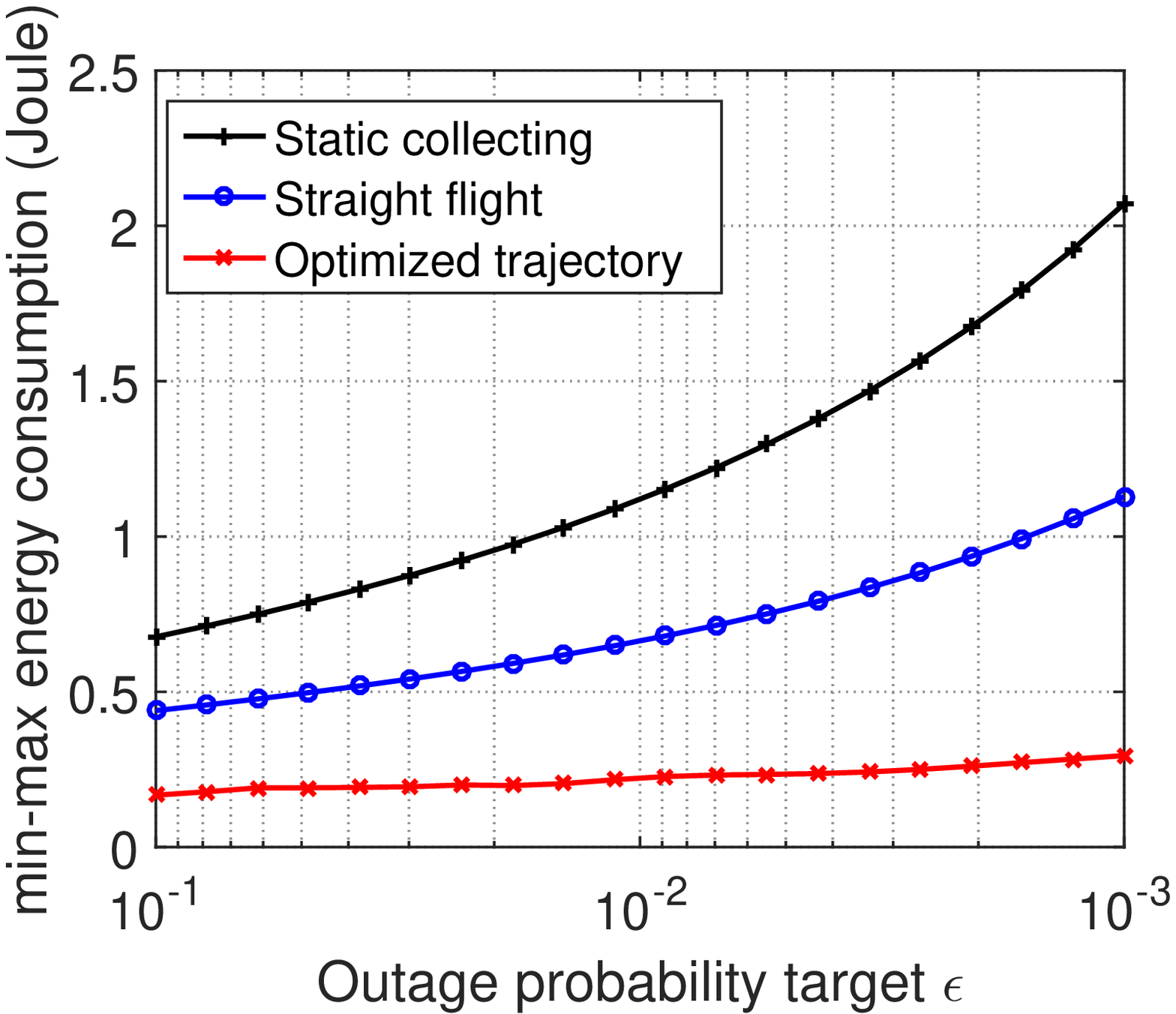}}
\end{minipage}}
\caption{Min-max energy consumption $\theta$ versus the sensing data size $S_k$ or outage probability target $\epsilon$.}
\label{fig:graph2}\vspace{-0.22in}
\end{figure}

In Fig. \ref{fig:graph2}, we compare the min-max energy consumption of our optimized trajectory with that of straight flight and the conventional static collecting scheme, where the data collector is deployed in the fixed location at the geometric center of all SNs. For fair comparisons, the wake-up schedule of the straight flight and static collecting schemes are also optimized by Algorithm 2. Fig. \ref{fig:graph2} shows the min-max energy consumption versus the sensing data size $S_k$ or outage probability target $\epsilon$ for $T=100$s. It is observed that our proposed trajectory design significantly outperforms the two benchmark schemes, and the performance gain is more pronounced as $S_k$ increases or $\epsilon$ decreases. This is expected since with our proposed scheme, the UAV can fly closer to or
even stays above the SNs for data collection with better channels, due to which the SNs can transmit at higher data rate reliably with less transmission time and thus save energy consumption.

\section{conclusion}
This letter proposes a novel design for
energy-efficient data collection in UAV-enabled WSNs. The SNs' wake-up schedule and UAV's trajectory are jointly optimized to minimize the maximum energy consumption of all SNs while ensuring reliable data collection in fading channels. With the successive convex optimization
technique, an efficient iterative algorithm is proposed to find a sub-optimal solution. Numerical results show significant energy savings with our proposed design as compared to the benchmark schemes.

\end{document}